# Knowledge Graphs Construction from Criminal Court Appeals: Insights from the French Cassation Court

Alexander V. Belikov[1]
Sacha Raoult[2,3]

**Despite growing interest, accurately and reliably representing unstructured data, such as court decisions, in a structured form, remains a challenge. Recent advancements in generative AI applied to language modeling enabled the transformation of text into knowledge graphs, unlocking new opportunities for analysis and modeling. This paper presents a framework for constructing knowledge graphs from appeals to the French Cassation Court. The framework includes a domain-specific ontology and a derived dataset, offering a foundation for structured legal data representation and analysis.**

## Introduction

Data gathering for quantitative sentencing research in criminal law has traditionally relied on three primary methodologies: analyzing government-maintained sentencing databases, manually coding data from paper case files, or completing observation sheets during courtroom proceedings. Each of these approaches has played a major role in shaping our understanding of judicial practices, yet they come with distinct limitations and advantages. Government databases offer a structured and often comprehensive overview of sentencing trends, but they are typically limited by the scope of their collection and the inflexibility of pre-coded categories. Manual coding of paper files allows for nuanced insights into case-specific factors, yet it is labor-intensive and time-consuming. Similarly, courtroom observation sheets capture real-time data about judicial proceedings but are constrained by the subjective interpretations of observers and the practical limitations of covering a broad spectrum of cases.

Recent advances in artificial intelligence (AI) present an opportunity to transform this landscape. Automated analysis of judicial decisions using AI offers the potential to unlock insights that traditional methods might overlook. By leveraging big data, this approach allows for the large-scale processing and analysis of judicial texts, making it possible to identify patterns, trends, and correlations across vast datasets. The French open-access judicial database Judilibre provides a unique opportunity to demonstrate the power of this approach. Through automated AI tools, decisions from Judilibre can be processed to extract statistical insights, offering a more dynamic, scalable, and objective way to analyze sentencing practices. This paper explores the use of AI to analyze judicial decisions in the context of criminal law, focusing on its ability to complement and surpass traditional methodologies by providing a big data-driven perspective on judicial practices.

[1] Growgraph, Paris; [2]Institut Universitaire de France [3]Aix-Marseille University;
Correspondence should be addressed to alexander@growgraph.dev or sacha.raoult@sciencespo.fr

Knowledge Graphs (KG) are the most generic and potentially accurate form of representation of knowledge. Although there is no universally accepted definition of what a knowledge graph is, it is usually understood as a collection of edges between well-defined nodes. This means that the nodes are grouped into types or classes formalized by domain experts. Edges, in this context known as relations or properties, may also belong to types and help define the structure of the knowledge being represented. The utility of KGs is manifold, it ranges from catalogization and document classification to retrieval-augemented generation (RAG) (2024, Colombo, A.), aiding chat agents in navigation in large corpora, and predictive analysis.

The choice of KGs as the representation format for judicial data stems from their unique ability to model complex relationships in a structured, interpretable, and flexible manner. Unlike plain text analytics, which often struggles with extracting meaningful insights from unstructured narratives, or tabular data, which imposes rigid schemas that can oversimplify nuanced information, knowledge graphs strike a balance between structure and expressiveness.

KGs enable the representation of entities (such as persons, events, or legal decisions) as nodes and their interconnections (such as relationships, causality, or chronology) as edges. This representation is inherently well-suited to the legal domain, where the interplay between various actors, events, and legal provisions is critical. For example, KGs can efficiently represent relationships like "Person A convicted for Crime B with Punishment C," while simultaneously linking the case to relevant statutes and precedent cases.

## Related Work

Recent advancements in knowledge graph construction from legal texts have built upon a variety of approaches, from rule-based systems to deep learning techniques. Early methods, such as those leveraging predefined templates and manually curated rules, were effective in structured datasets but lacked adaptability and scalability (LegalIE, 2013). The advent of machine learning introduced supervised models like Conditional Random Fields (CRFs) and Support Vector Machines (SVMs), which improved automation but required extensive feature engineering and annotated datasets (Rodrigues et al., 2019). More recently, deep learning models such as LSTMs and attention-based transformers have enhanced tasks like named entity recognition and relation extraction, enabling more accurate and scalable solutions (Zhou et al., 2024). While BERT-based architectures now dominate legal NLP, alternative methods such as graph neural networks (GNNs) and hybrid systems combining statistical and rule-based models remain important, particularly in domains with sparse labeled data. Ontologies like LegalRuleML and LKIF have provided foundational structures for legal knowledge representation, but they often lack the granularity or domain specificity required for criminal law (Palmirani et al, 2011, Leone et al., 2018). This paper builds upon these efforts by proposing a tailored criminal ontology that integrates elements of existing frameworks while addressing the nuances of criminal law, offering a more robust approach to distilling knowledge from legal texts.

## Methodology

### Data Acquisition and Preprocessing

We work with appeals before the Supreme court for criminal cases in France (https://www.courdecassation.fr/recherche-judilibre). Each document of 6-8 pages is a formal court decision with respect to an appeal, so it contains such information as the description of a crime, the status of a convict, the punishment, as well as references to the previous court decisions. Court decisions are anonymized. The data is available in pdf format.

### Approaches

Knowledge graph generation from text is a complex task. Until very recently one would have to establish and combine pipelines for named entity recognition (to discern the types of candidates, e.g. Person vs Location), relation extraction (to identify the modality and the direction of relation between the subject and the object, e.g. A causes B), co-referencing (to link candidates referring to the same concept) and entity linking (to map proper nouns to well-known entities, such names of cities). With the advent of generative Large Language Models (LLMs) of GPT type, it became possible to unify these tasks. Instead of developing task-specific models it is now possible to design task-specific LLM prompts to induce the generation of knowledge graph constituents and receive formatted output in a robust manner.
Drawn by the setup simplicity (one general model vs an ensemble of models) as well as undeniable success of GPT-type models we opted for experiments with the class of generative LLMs. It is well known that language models are designed to perform well in specific domains and are not guaranteed to operate as well in domains different from which they were trained on. We considered two approaches corresponding to the property graph and RDF triple store representation options.

### Property Graph

A python package *llama_index*, designed as a data framework for LLM application, provides a utility of property graph generation (**SchemaLLMPathExtractor**). It uses an LLM to interpret the input and an embedding model to match input data to entities. As an input it requires a list of possible entity types, a list of possible relations and a schema to validate what entity types are admissible for specific relations. *llama_index* provides the utility of directly ingesting the resultant knowledge graph into **Neo4j**, a popular graph database.

### RDF Triple representation

As an alternative approach we query LLM directly with a request to generate an RDF structured output in *turtle* (**Terse RDF Triple Language**) format. In order to normalize the generation of triples we developed a RDF criminal ontology (incorporated into the prompt). In this manner we assure that the LLM is focusing in the classes and properties present in the ontology. Due to the stochastic nature of generative LLMs, even though our prompt contains an instruction to provide input in *turtle* format, which may require formatted (typed) inputs such as positive integers, or timestamp in ISO format, from time to time the LLM generates erroneous output that does not conform to declared format. Since the RDF framework is focused on standardization and interoperability, we are using FOAF, RDFS and schema.org ontologies, where appropriate.

The choice between property graph models and RDF triples in knowledge graph construction involves several trade-offs that extend beyond accuracy. Property graphs, such as those implemented in Neo4j, excel in ease of querying due to their intuitive, developer-friendly interfaces and support for flexible schemas. They are particularly suited for exploratory data analysis and applications requiring dynamic updates, as their schema-on-read approach allows for modifications without extensive re-engineering. However, property graphs often lack standardization, making interoperability between systems a challenge. In contrast, RDF triples adhere to established standards like W3C's RDF framework, enabling seamless data integration and sharing across different platforms. RDF triples also leverage ontologies to enforce semantic consistency, which is essential for interoperability in complex domains such as law. Despite these advantages, querying RDF data can be less intuitive, often requiring familiarity with SPARQL, and the schema-on-write approach can complicate dynamic graph modifications. These trade-offs highlight the need to align the choice of representation with the specific requirements of the legal application, such as scalability, cross-system integration, or ease of maintenance.

## Knowledge Graph Construction

### Ontology Design

Starting from the 1990s a multitude of ontologies were developed at the cross-section of Law and information representation (2019, Rodrigues et al). The deluge inspired even legal ontology search tools (2018 Leone et al.). With a focus on a specific domain we created our version of criminal ontology `` `@prefix fca: <https://growgraph.dev/fcaont#> .` ``, elucidating the details of the judicial process, where a case may receive an appeal, relations between the type Person, that may have roles Convict and Victim, events; conviction and punishment and its specifics, such the duration of custodial punishments and the amount of monetary punishments, types of appeal decisions etc.

The ontology was developed in a semi-automated way, via interaction with the **GPT-4o mini** and **Claude 3.5 Sonnet** models. We iterated starting from the first version, requesting the model to add new types and relations, all the while making sure the generated output is standardized, corresponds to our expectations and covers certain edge cases, such as:

- Introducing relevant properties. For example, if class **Location** is present in the ontology and class **Crime** has **Location**, then the model may identify the court location as a crime location. The model tends to project specific entities onto entities provided in the ontology, given available properties

- Some literals may have constraints and/or preferred representation, such non-negative integers for age in years, duration and date according to ISO 8601, currency codes. While in general **GPT-4o mini** is very good at generating, for example, duration in ISO 8601, sometimes it fails. In order to improve the accuracy we include an explicit instruction in the prompt to validate such outputs.

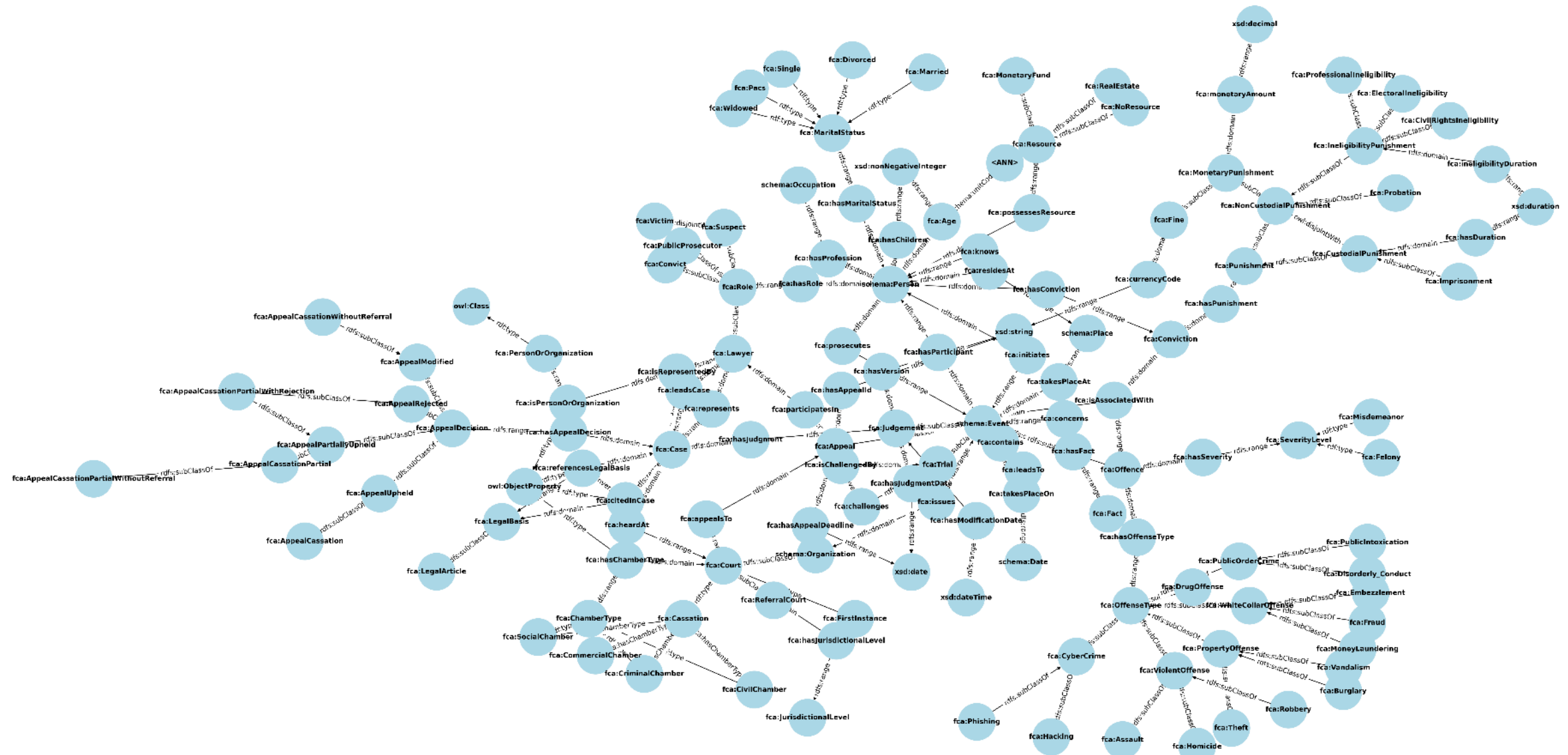

**Fig. 1.** Criminal ontology (simplified for clarity).

Once we obtained a satisfactory ontology (from the point of view of experts), we included it into the prompt for each analyzed appeal.

We evaluated the property graph approach accuracy in the ballpark of 50-60% while the triple approach results a priori demonstrated > 90% accuracy and hence we concentrated on the latter case.

## Evaluation

In order to evaluate the performance of different models we created a ground truth dataset based on 20 appeals (Cassation Cour Appeal ***CCP2023*** *dataset)*. We fed only the selected excerpts containing the information of interest from the reference dataset and validated the derived knowledge graphs with the help of experts. Running the tests on CCP dataset we conclude that out-of-the-box precision of **GPT-4o mini** is 93%, while the recall is at 89%.

## Implementation

We used *langchain* python package to query LLMs[2]. Our query is composed of the ontology and specific guidance rules. We used *pymupdf* python package to parse pdfs. The query result contains a turtle formatted document, which we validated using *rdflib* python package, which is then placed into a Fuseki triple store. We ran our experiments with the **GPT-4o mini**. Preliminary experiments excluded **llama-3.1 7B**, due to the low quality of the output. We estimate the processing cost at 2.5$ per 1K documents as of Dec 2024. Our prompt included a suggestion to LLM to provide comments and suggestions, which can serve as the basis for the ontology improvement. We note that there is an interplay between the size, granularity of the ontology and the quality of KG generation (and the cost). Larger ontologies provide better guidance to the LLM, but require submitting a greater number of tokens, affecting the cost. At the same time a prompt (with ontolog)  that is too large might surpass the short memory capacity, leading the model to forget part of important information, which is an issue attracting some attention (2024, Z. Wang et al).

## Results

We process 2820 appeals to the criminal chamber of the Supreme court from 2023. The dataset is available at On average 30 triples are generated per appeal (cf. Fig. 2, left for the distribution of the number of triples). Appeals cover various types of offense, with violent offense being by far the most popular, cf. Fig. 2, right. As a step towards preliminary analysis, we plotted the distributions of punishment duration and fine amount per appeal decision, cf. Fig.4. Our preliminary analysis shows that punishment duration and fine amount have negligible effect on the decision of the Cassation Court.

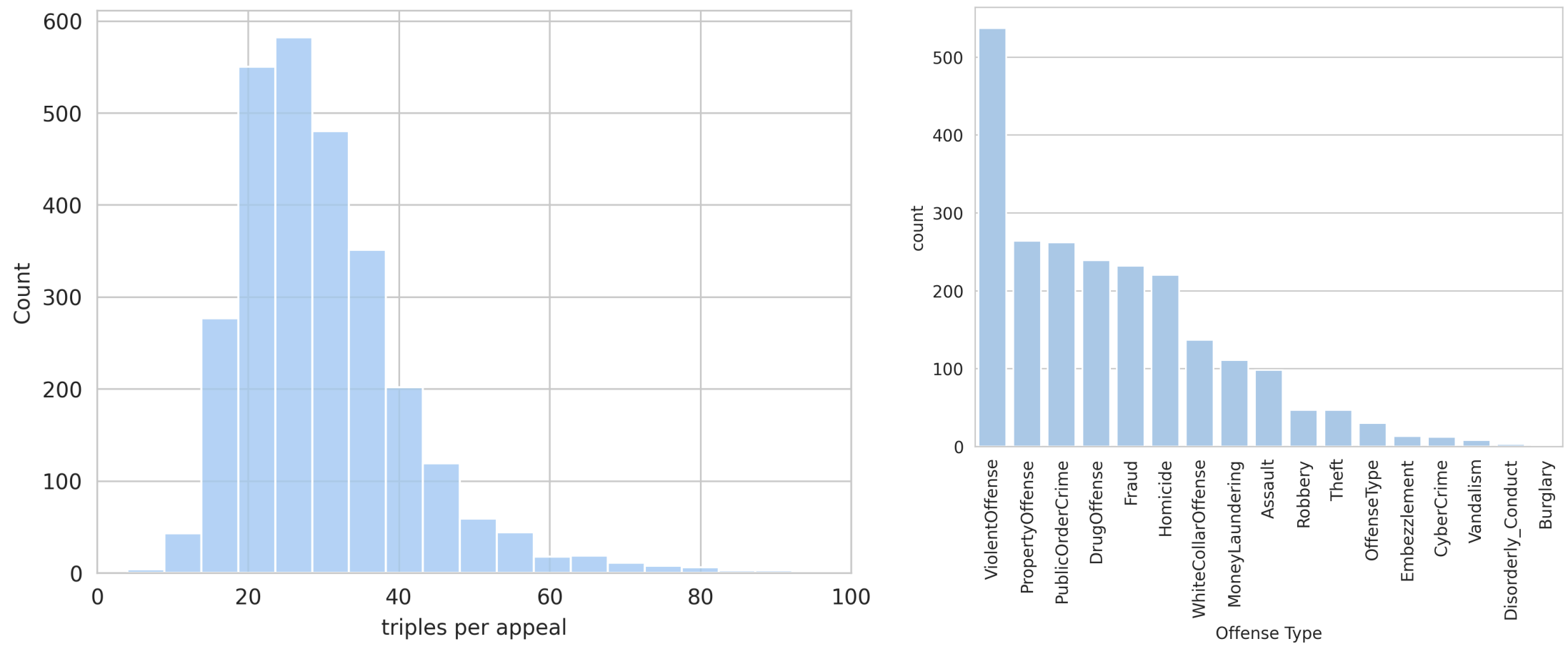

**Fig. 2.** Left: the distribution of triple counts per appeal. Right: the distribution of crimes by the type of offense.

[2] The codes are available at https://github.com/growgraph/legal_ie

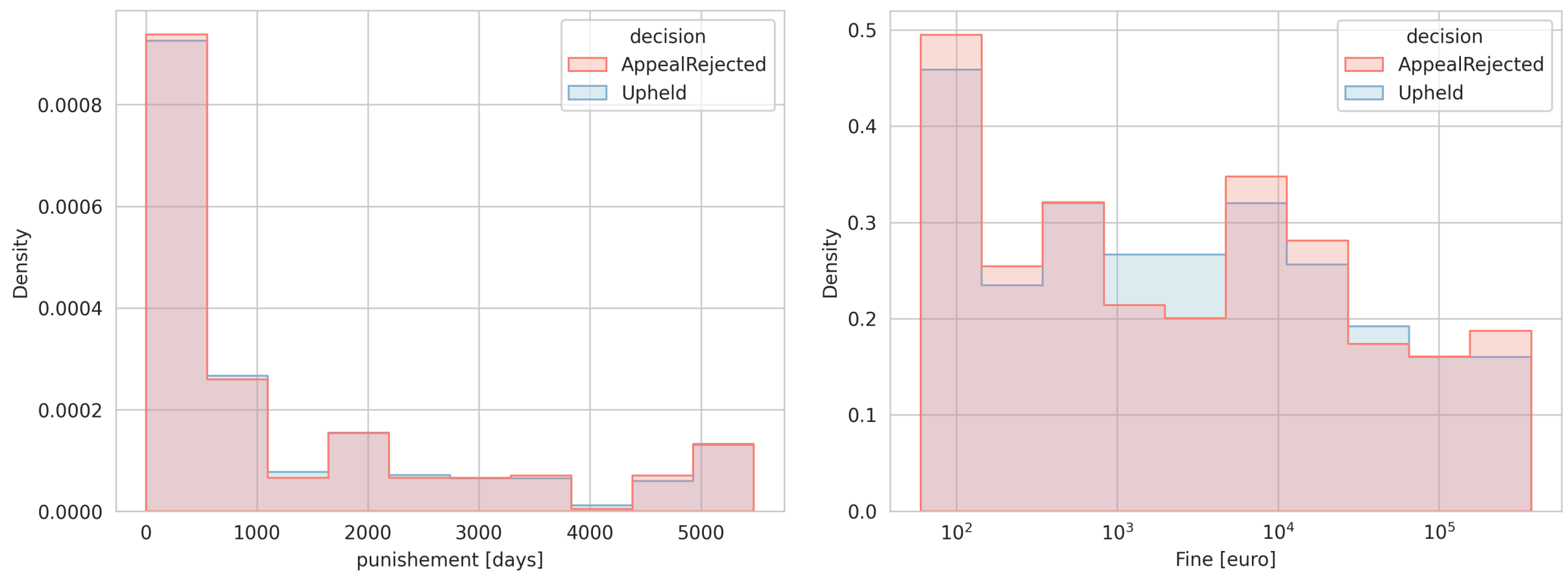

**Fig. 3.** Left: the distribution of duration of custodial punishments per appeal in days. Right: the distribution fine amounts by appeal in euros. Red denotes rejected appeals, while blue - upheld. In both figures the distributions are normalized with respect to the group.

## Conclusions and Future Work

In this work we developed an end-to-end system for information extraction from legal documents from the Supreme Court in France by knowledge graph distillation. In particular we demonstrated the power of iterative LLM-aided approach of ontology development, and evaluated the efficiency of ontology based KG generation in the legal domain.

The quality of constructed knowledge graphs depends on the architecture, training procedure (training strategies, size and specialization of the training dataset) of the LLM in question as well as on the such qualities of ontology, as self-consistency, alignment with the domain of interest and granularity.

The improvements on the KG generation performance include:

- using an improved ontology
- pretraining the LLM using domain specific labeled datasets
- providing the ontology in a vector store

Our methodology can facilitate such applications as retrieval augmented generation for user interaction, deduction of facts, statistical and predictive data analysis.